\documentstyle[12pt]{article}

\def\be{\begin{equation}}
\def\ee{\end{equation}}
\def\a{\alpha}
\def\ra{\rangle}
\def\la{\langle}
\def\tg{\mbox{tg}} 
\def\sh{\mbox{sh}}

\def\tp{t^{\prime}}

\def\w{\omega} 
\def\l{\lambda}
\def\L{\Lambda}

\begin{document}
 
\begin{center}
{\bf Threshold singularities in the XXZ- spin chain.}
\end{center}
\vspace{0.2in}
\begin{center}

{\large A.A.Ovchinnikov}

\end{center}   

\begin{center}
{\it Institute for Nuclear Research, RAS, Moscow}
\end{center}   
 
\vspace{0.2in}

\begin{abstract}

We calculate the critical exponents of the threshold 
singularity for the spectral density of the XXZ- spin chain 
at zero magnetic field for the lower threshold. 
We show that the corresponding phase shifts  
are momentum - independent and coincide with predictions 
of the effective mobile impurity Hamiltonian approach. 
We show that for the eigenstates with the high-energy 
particle with the momentum $k$ which is much larger than 
the momenta of the particle-hole excitations the formfactor 
is not described in the framework of the Luttinger model 
even in the limit $k\ll p_F$ but should be evaluated in the 
framework of the effective mobile impurity Hamiltonian 
approach.

\end{abstract}

\vspace{0.2in}

{\bf 1. Introduction}

\vspace{0.2in}

The XXZ- spin chain is interesting both from the theoretical 
and the experimental points of view. 
The Hamiltonian of the model has the form: 
\be
H=\sum_{i=1}^{L}(S_{i}^{x}S_{i+1}^{x}+S_{i}^{y}S_{i+1}^{y}
+\Delta S_{i}^{z}S_{i+1}^{z})+h\sum_{i}S^{z}_i, 
\label{model}
\ee
where the periodic boundary conditions are assumed, $L$ is the 
length of the chain, the anisotropy parameter $\Delta=\cos(\eta)$ 
and $h$ is the magnetic field. 
Recently along with the calculation of the asymptotics of the 
correlators for the XXZ- spin chain and the other integrable models 
\cite{LP},\cite{H}, the calculation of the critical exponents 
of the threshold singularities of different correlators 
attracted much interest. The first attempts \cite{PWA},\cite{CP} 
to calculate these exponents for the model (\ref{model}) 
are based on the notion of the effective mobile impurity Hamiltonian 
\cite{G},\cite{RMP} combined with the calculation of the parameters 
with the help of the Bethe Ansatz. However the predictions of this 
approach contradict the universal behaviour of the phase shifts 
at small momentum $k\rightarrow 0$ \cite{S}.  
Thus there is an obvious contradiction between the results of 
ref.\cite{PWA},\cite{CP} and the predictions of the Luttinger liquid 
Bosonization approach \cite{S},\cite{O} at low momentum.   
The formfactors of the exactly solvable models  were 
studied in the framework of the rigorous approach \cite{K1}. 
Although in general the formfactors of the XXZ- spin chain 
corresponding to the long-distance behaviour of the 
correlation functions were obtained in \cite{K1}, and the 
singularities for the 1D Bose gas were obtained in \cite{K2} 
using the formfactor approach, the clear results for the 
singularities of the XXZ- spin chain at zero magnetic field 
are still absent. Also in \cite{K1},\cite{K2} all the critical 
exponents are expressed through the two-particle dressed 
phase shifts \cite{Korepin}. However the calculation of these 
momentum-dependent phase shifts in the XXZ- spin chain and comparison 
of the results with the predictions of ref.\cite{PWA},\cite{CP} 
is a separate problem.

It is the goal of the present paper to obtain the critical 
exponent of the singularity of the correlator 
$\la S^{+}_xS^{-}_0\ra$  at the lower threshold in the model 
(\ref{model}) at $h=0$ with the help of the rigorous approach 
\cite{K1},\cite{K2} and compare the results with the predictions of 
of ref.\cite{PWA},\cite{CP} and the universal phase shifts \cite{S}. 
Our results confirm the phase shifts \cite{PWA} for the XXZ- spin chain 
at zero magnetic field and contradict the predictions of 
ref.\cite{S},\cite{O} at small momentum. 
These results imply that in general the local spin operator 
cannot be represented as the exponential operator in the effective 
Luttinger liquid model even at the low energies. 
At the momentum of the high-energy particle (hole) much larger 
than the momenta of the low-energy particle-hole excitations the 
formfactors should be calculated in the framework of the effective 
mobile impurity Hamiltonian approach.

\vspace{0.2in}

{\bf 2. Generalized Cauchy determinant at finite magnetic field.}

\vspace{0.2in}

Let us consider first the particular case of the XX- spin chain ($\Delta=0$). 
The solution of the model has the following form \cite{LSM}.   
We assume for simplicity $L$- to be even and 
$M=L/2$ to be odd ($S^z=0$ for the ground state and $(M-1)$- even, we also 
assume $L$ to be even so that the ground state is not degenerate). 
Then each eigenstate in the sector with $M$ particles (up-spins) is 
characterized by the set of the momenta $\{p\}=\{p_1,\ldots p_M\}$ such that 
$p_i=2\pi n_i/L$, $n_i\in Z$ and each eigenstate in the sector with 
$M-1$ particles is characterized by the set of the momenta 
$\{q\}=\{q_1,\ldots q_{M-1}\}$, $q_i=2\pi(n_i+1/2)/L$, $n_i\in Z$. 
The ground-state in the sector with $M$ particles (up-spins) is given by the 
configuration $\{p\}=\{p_1,\ldots p_M\}$, $p_i=2\pi/L(i-(M+1)/2)$, ($M$- is odd), 
and the ground state in the sector with $M-1$ particles 
$\{q_0\}=\{q_1,\ldots q_{M-1}\}$, $q^{(0)}_i=2\pi/L(i-M/2)$. 
Equivalently one can take the shifted momenta 
\[
p_i=(2\pi/L)(i),~~i=1,\ldots M, ~~~~q^{(0)}_j=(2\pi/L)(j+1/2), ~~j=1,\ldots M-1. 
\]
In terms of the sets of the momenta $\{p\}$ and $\{q\}$ the 
formfactor $\la\{q\}|S_0^{-}|\{p\}\ra$ can be represented in 
the following form \cite{OFF}: 
\be
\la\{q\}|S_0^{-}|\{p\}\ra\sim
\frac{\prod_{i<j}\sin((p_i-p_j)/2)\prod_{i<j}\sin((q_i-q_j)/2)}{\prod_{i,j}
\sin((p_i-q_j)/2)}. 
\label{xx}
\ee
Let us note that the expression (\ref{xx}) for the formfactor is exact   
for an arbitrary state characterized by the momenta $q_1,\ldots q_{M-1}$.

Now let us consider the general case of the XXZ- spin chain in the 
finite magnetic field $h\neq0$. Let $\{t\}=t_1,\ldots t_M$ and 
$\{\l\}=\l_1,\ldots\l_{M-1}$ to be the spectral parameters (rapidities) 
of the ground state $|\{t\}\ra$ and the excited state $|\{\l\}\ra$, where the 
number of roots $M$ is connected with the magnetic field $h$. 
We are interested in the formfactor of the local operator $S^{-}_0$ 
of the form $\la\{\l\}|S^{-}_0|\{t\}\ra$. Suppose the state $|\{\l\}\ra$ 
contains the hole with the rapidity $t_0$ corresponding to the 
momentum $k$. Then the part of the formfactor which contains the 
information about the low-energy particle-hole excitations in the 
state $|\{\l\}\ra$ has the form of the generalized Cauchy determinant 
\be
\la\{\l\}|S^{-}_0|\{t\}\ra
\sim\frac{\prod_{i<j}\sh(t_i-t_j)\prod_{i<j}\sh(\l_i-\l_j)}
{\prod_{i,j}\sh(t_i-\l_j)}. 
\label{cauchy}
\ee 
The other factors depend on $k$ but not on the quantum numbers of the 
low-energy excitations in the state $|\l\ra$. One can present the 
following arguments in favor of (\ref{cauchy}). First, one can see that 
in the XX- limit the determinant (\ref{cauchy}) reproduce the exact result 
(\ref{xx}) which follows from the formula $e^{\pi t/\eta}=\tg(p/2+\pi/4)$, 
which connects the rapidity $t$ with the corresponding momentum $p$. 
Second, this statement was proved by means of the complicated analysis 
in \cite{K1}. Third, in fact the results obtained from (\ref{cauchy}) 
do not depend on the specific form of the function which enters 
(\ref{cauchy}), so this equation can be considered as a natural 
hypothesis confirmed by the example of the XX- spin chain. 
Let us stress that we can use the expression (\ref{cauchy}) only at 
$h\neq0$, while the corresponding expression at $h=0$ is not known. 
Thus in the present paper we will perform all calculations for $h\neq0$ 
and take the limit $h\rightarrow 0$ only in the final expressions for 
the phase shifts.

\vspace{0.2in}

{\bf 3. Singularity at the lower threshold.}

\vspace{0.2in}

Here we calculate the phase shifts for the XXZ- spin chain at $h=0$ 
starting from the expression (\ref{cauchy}) for the formfactors. 
We perform the calculations at $h\neq0$, or at finite cutoff $\L$ 
for the rapidities $t_{\a}$, $t_{\a}\in(-\L,\L)$, and take the limit 
$\L\to\infty$ only at the end of the calculations. 
We consider the configuration $\{\l\}$ obtained from the vacuum configuration 
of $M-1$ roots by removing the single root at the position $t_0$ (hole) 
and adding an extra root $\l_M$ at the right end of the interval $(-\L,\L)$. 
Near the right end of this interval the values of the roots and their 
difference have the form: 
\be
t_i\simeq t_M-\frac{1}{LR(\L)}i, ~~~~\l_j\simeq \l_M-\frac{1}{LR(\L)}j, ~~~~
t_i-\l_j\simeq\frac{1}{LR(\L)}\left(i-j+LR(\L)(t_M-\l_M)\right), 
\label{roots}
\ee
where $t_M$ and $\l_M$ are the maximal roots and $R(t)$ is the density 
of roots at $t\in(-\L,\L)$. 
The calculation of the Cauchy determinant (\ref{cauchy}) corresponding 
to a given configuration of particles and holes at the right (left) 
Fermi- points is quite simple and gives the known expression 
\cite{K1},\cite{Shashi},\cite{Opla} with the phase shift parameter: 
\be 
\delta_1=LR(\L)(t_M-\l_M)=RW(\L), 
\label{delta}
\ee
where the function $W(t)$ is defined as $W(t_{\a})=L(t_{\a}-\l_{\a})$, 
where $t_{\a}$ and $\l_{\a}$ - are the corresponding roots from the sets 
$\{t\}$ and $\{\l\}$ starting from the roots $t_M$, $\l_M$. 
The equation for the function $RW(t)=R(t)W(t)$ 
is obtained by the subtraction 
of the corresponding Bethe Ansatz equations for  $t_{\a}$ and $\l_{\a}$ 
and takes the form: 
\be 
2\pi RW(t)+\int_{-\L}^{\L}d\tp\phi_2^{\prime}(t-\tp)RW(\tp)= 
-\pi+\phi_2(t-t_0), 
\label{1}
\ee
where the function $\phi_2(t)=-(1/i)\ln((t-i\eta)/(t+i\eta))$. 
Clearly, for the vacuum case ($\{\l\}$- does not contain a hole) 
we obtain the same equation (\ref{1}) with the right-hand side equal to 
$\pi+\phi_2(t+\L)$. 
The solution of the equation (\ref{1}) is quite standard and is given 
by the following simple formulas. 
First, for the term $-\pi$ at the right-hand side of (\ref{1}) the solution 
is given by the dressed charge function $Z(t)$ defined by the equation 
\[
2\pi Z(t)+\int_{-\L}^{\L}d\tp\phi_2^{\prime}(t-\tp)Z(\tp)=2\pi.   
\]
It is known that $Z(\L)=1/\sqrt{\xi}$, where $\xi=2(\pi-\eta)/\pi$ is 
the standard Luttinger liquid parameter for the XXZ- spin chain. 
Second, to obtain the solution for the second term $RW_1(t)$,  
for the right-hand side, we rewrite the equation (\ref{1}) 
in the form: 
\be 
\chi(t)=f(t)+\int_{0}^{\infty}d\tp F(t-\tp)\chi(\tp), 
\label{chi}
\ee
where $\chi(t)=RW_1(t+\L)$, the Fourier transform 
$F(\w)=\phi_2^{\prime}(\w)/(2\pi+\phi_2^{\prime}(\w))$ 
and the function 
$f(t)=\tilde{F}(t+\L-t_0)$, where the function $\tilde{F}(t)$  
is defined by the equation $\tilde{F}^{\prime}(t)=F(t)$. 
The general solution of the equation (\ref{chi}) is 
\be 
\chi^{+}(\w)=G^{+}(\w)\int\frac{d\w^{\prime}}{2\pi i}
\frac{1}{(\w^{\prime}-\w-i0)}G^{-}(\w^{\prime})f(\w^{\prime}), 
\label{general}
\ee
where $\chi^{+}(\w)=\int_0^{\infty}dte^{i\w t}\chi(t)$, the functions 
$G^{\pm}(\w)$ holomorphic at the upper (lower) half-plane of the variable 
$\w$ are defined by the equation $F(\w)=1-1/G^{+}(\w)G^{-}(\w)$ 
(for example, see \cite{YY}) and in our case 
$f(\w)=e^{-i\w(\L-t_0)}\tilde{F}(\w)$. 
Taking the limit $\w\to\infty$ in the equation (\ref{general}) 
we obtain the contribution to the phase shift: 
\be
RW_1(\L)=\chi(0)=\int\frac{d\w}{2\pi}e^{-i\w(\L-t_0)}G^{-}(\w)
\tilde{F}(\w). 
\label{contrib}
\ee
To obtain the Fourier transform $\tilde{F}(\w)$ one can use the 
following identity: 
\[
\phi_2(\w)=\frac{i}{\w+i}\phi_2^{\prime}(\w)-2\pi(\pi-2\eta)\delta(\w). 
\]
Thus the corresponding contribution to the phase shift (\ref{contrib}) 
is found.  
To calculate (\ref{contrib}) at $(\L-t_0)\gg 1$ one should consider the 
integration contour at the lower half-plane of the complex variable $\w$. 
The leading  $\sim O(1)$ is given by the residue of the pole at $\w=-i0$. 
Thus taking into account the contribution of the first term in (\ref{1}) 
we obtain the phase shift for the right Fermi-point at $(\L-t_0)\gg 1$: 
\be 
\delta_1=\frac{1}{2\sqrt{\xi}}+\frac{1}{2}\sqrt{\xi}(1-1/\xi)=
\frac{\sqrt{\xi}}{2}-\frac{1}{\sqrt{\xi}}. 
\label{pwa} 
\ee
At the same time at the left Fermi-point the phase shift $\delta_2$ 
is equal to its ``vacuum'' value $\delta_2=\sqrt{\xi}/2$. 

One can easily verify that these values of the phase shifts 
$\delta_1$, $\delta_2$ coincide with the predictions of the 
mobile impurity Hamiltonian method \cite{PWA}. 
At the same time at $(\L-t_0)\gg 1$ the result for $\delta_1$ does not 
coincides with the prediction of the Luttinger liquid Bosonization 
approach $\delta_1=1-\sqrt{\xi}/2$. 

In the opposite limit $(\L-t_0)\ll 1$ from the equation (\ref{contrib}) 
we obtain exactly this universal value $\delta_1=1-\sqrt{\xi}/2$. 
However since in this limit $(\L-t_0)\simeq k/2\pi R(\L)$ where $k$- 
is the momentum of the hole the universal phase shift is reproduced 
only at the very small momentum $k\ll R(\L)\sim e^{-\pi\L/\eta}$ 
(note that $\L\to\infty$). 
Using the formfactors one can easily calculate the threshold singularity 
for the dynamical structure factor \cite{O}: 
\[
A(\w,k)=\sum_x\int dte^{i\w t-ikx}\la S^{+}_x(t)S^{-}_0(0)\ra\sim 
\frac{1}{(\w-\epsilon(k))^{\mu}}, 
\]
where $\epsilon(k)$- is the known excitation energy of the single 
hole (particle) and the critical exponent is given by the expression: 
\be
\mu=1-\delta_1^2-\delta_2^2=2-1/\xi+\xi/2.   
\label{mu} 
\ee
The reason for the sharp transition from the Luttinger liquid result 
for $\delta_1$ to the momentum-independent results of ref.\cite{PWA} 
is as follows. If $t_0$ is sufficiently close to to the extra root 
$\l_M$ (or $\L$) there is no shift of the roots between $t_0$ and $\L$ 
and clearly the phase shift $\delta_1=1-\sqrt{\xi}/2$, where 
$\sqrt{\xi}/2$ is the ``vacuum'' value of the phase shift. 
That is what one expects in the framework of the Luttinger liquid. 
However at $(\L-t_0)\gg 1$ the roots between $t_0$ and $\L$ acquire
the additional shift to the left, which is described by the by the 
equation (\ref{1}), which makes the naive predictions of the Luttinger 
liquid (Bosonization) incorrect. 
At the same time in this case the mobile impurity Hamiltonian approach 
is still applicable for the formfactors and leads to the results for 
$\delta_1$, $\delta_2$ in agreement with the rigorous approach 
\cite{K1},\cite{K2} (see (\ref{cauchy})). 
The main result of the present letter is that contrary to the naive 
expectations this shift of the roots is model-dependent and cannot be 
described in the framework of the Luttinger liquid. 
Note that the main assumption of mobile impurity Hamiltonian approach 
is the identification of the Jordan-Wigner fermion $\psi$ with the 
impurity operator $d$, $\psi^{+}\to d^{+}$. However the similar 
substitution $\psi^{+}\to a^{+}$, where $a^{+}$- is the Luttinger liquid 
fermionic operator, is made in the framework of the Bosonization 
approach.  
Summarizing, we have found that for the XXZ- spin chain the matrix element 
$\la p,low|S^{-}|0\ra$, 
where $p$ is the momentum of the hole cannot be calculated in the regime 
when the momentum $p$ is much larger than the particle-hole momenta 
$p_i$, $q_i$ in the state $\la low|$ (but still much smaller than the 
Fermi- momentum $\sim p_F$) as the matrix element of the exponential 
operator 
\[
\la p,low|e^{i\pi\sqrt{\xi}(\hat{N}_1-\hat{N}_2)}|0\ra, 
\]
where $\hat{N}_1$, $\hat{N}_2$ are the rescaled Luttinger liquid fields. 
Rather in this limit one should use the effective mobile impurity 
Hamiltonian to evaluate this formfactor correctly. 
The reason for this surprising result is not clear at present time. 
We describe the correct application of this method for the XXZ- spin chain 
in the next Section. 
However let us note that the long-distance asymptotics of the 
correlators are reproduced correctly both in the framework of the 
Bosonization approach and from the formfactors (\ref{cauchy}).

\vspace{0.2in}

{\bf 4. Mobile impurity Hamiltonian.}

\vspace{0.2in}

Here we briefly describe the application of the effective mobile 
impurity Hamiltonian for the XXZ- spin chain. Although the parameters 
of the Hamiltonian ($V_1$, $V_2$, see below) are differ from the results 
of ref.\cite{PWA}, the final results for the phase shifts are exactly 
the same as in ref.\cite{PWA}. 
The effective mobile impurity Hamiltonian has the form: 
\be
H=H_{LT}+d^{+}(\epsilon(k)-iv_{d}\partial_x)d(x)+(V_{1}n_1+V_{2}n_2)n_d, 
\label{Heff}
\ee
where $d$ ($d^{+}$) are the operators corresponding to the high-energy 
hole with the $k$- dependent velocity 
$v_d=\partial\epsilon(k)/\partial k$, $n_d=d^{+}d$, $n_{1,2}(x)$ are the 
standard low-energy particle densities and 
\be
H_{LT}=\frac{\pi v}{2}\left(\xi(n_1+n_2)^2+(1/\xi)(n_1-n_2)^2\right) 
\label{LT}
\ee
is the Luttinger liquid Hamiltonian characterized by the standard 
Luttinger liquid parameter $\xi$ ($\xi=1/K$). The Hamiltonian 
(\ref{Heff}) can be diagonalized with the help of the unitary operator 
\be
U=\exp\left(i2\pi\int dy(\delta_{1}\hat{N}_1(y)+\delta_{2}\hat{N}_2(y))
n_d(y)\right), 
\label{U}
\ee
where $\partial_{x}\hat{N}_{1,2}(x)=\hat{n}_{1,2}(x)$ -  
are the rescaled fields $n_{1,2}$ (see (\ref{LT})) and $\delta_{1,2}$ 
are the phase shifts. The interaction terms in (\ref{Heff}) 
are cancelled if the phase shifts are connected with the parameters 
of the Hamiltonian $V_{1,2}$ in the following way \cite{G}: 
\be 
(V_1-V_2)\sqrt{\xi}=-\delta_2(v_d+v)+\delta_1(v_d-v), 
\label{G}
\ee
\[
(V_1+V_2)(1/\sqrt{\xi})=-\delta_2(v_d+v)-\delta_1(v_d-v). 
\]
The correct values of the parameters $V_1$, $V_2$ for the XXZ- spin 
chain were calculated in ref.\cite{O19}. The result has the form: 
\be 
V_1=-\pi v(1-K)\xi+\pi v_d(1-K), ~~~V_2=-\pi v(1-K)\xi-\pi v_d(1-K), 
\label{V}
\ee
where the second terms $\sim v_d$ correspond to the shift of the 
energy of the high-energy hole due to the density perturbations at 
the Fermi-points $n_{1,2}$. Note that this terms are absent in 
ref.\cite{PWA}. Thus we obtained the momentum-dependent parameters 
$V_1$, $V_2$. Substituting the parameters (\ref{V}) into the 
equations (\ref{G}) we obtain the phase shifts $\delta_{1,2}$ in 
(\ref{U}) which coincide with the phase shifts $\delta_{1,2}^{PWA}$ 
found in ref.\cite{PWA}: 
\be
\delta_1^{PWA}=-\frac{1}{2}\left(\sqrt{\xi}-1/\sqrt{\xi}\right), ~~~~
\delta_2^{PWA}=\frac{1}{2}\left(\sqrt{\xi}-1/\sqrt{\xi}\right).  
\label{PWA}
\ee 
Thus taking into account the presence of the Jordan-Wigner string 
we obtain the critical exponent in agreement with the expression 
(\ref{mu}) (and our expressions for the phase shifts $\delta_{1,2}$ 
coincide with the expressions $\delta_{1,2}^{PWA}$ (\ref{PWA}) 
after the Jordan-Wigner string is taken into account). 
The reason why the final result coincides with that of 
ref.\cite{PWA} is that the zero modes which come from the term 
$-iv_{d}d^{+}\partial_{x}d$ after the unitary rotation do not taken 
into account in \cite{PWA} (see ref.\cite{PWA1}).

\vspace{0.2in}

{\bf 5. Conclusion.}

\vspace{0.2in}

In conclusion, we obtained the critical exponent of the threshold 
singularity for the dynamical structure factor in the XXZ- spin chain 
at zero external magnetic field with the help of the rigorous method. 
The method is based on the expressions for the formfactors in the 
form of the generalized Cauchy determinants (\ref{cauchy}). 
Our results are in agreement with the predictions of the effective 
mobile impurity Hamiltonian method \cite{PWA},\cite{G} and contradict 
the naive Luttinger liquid theory predictions. 
Few remarks are in order here.  
The correct way to calculate the parameters of the effective 
mobile impurity Hamiltonian $V_1$, $V_2$ ($V_L$, $V_R$ in the notations 
of ref.\cite{G}) is to use the results of ref.\cite{O19}, where the 
interaction energy of the particle-hole pair in the XXZ- spin 
chain was calculated. This leads to the values of $V_1$, $V_2$ 
which are different from the values found in \cite{PWA} 
(the influence of the left and the right densities on the energy 
of the high-energy particle does not taken into account). 
However due to the other mistakes (see \cite{PWA1}) the final results 
for the phase shifts are correct. 
Finally, we found that the naive application of the Bosonization 
technique to the calculation of the formfactors does not lead to the 
correct results even at small energies. At the same time the 
application of the effective mobile impurity Hamiltonian method gives 
the correct momentum-independent results for the phase shifts and 
the critical exponent for the lower threshold singularity of 
the spectral density. 
Our main result is that in general the local spin operator 
cannot be represented as the exponential operator in the effective 
Luttinger liquid model even at the low energies.


\vspace{0.2in}  


\begin{thebibliography}{99}

\bibitem{LP}
A.Luther, I.Peschel, Phys.Rev.B9 (1974) 2911. 

\bibitem{H}
F.D.M.Haldane, Phys.Rev.Lett.47 (1981) 1840; J.Phys.C 14 (1981) 2585. 

\bibitem{PWA}
R.G.Pereira, S.R.White, I.Affleck, Phys.Rev.Lett. 100 (2008) 027206.

\bibitem{CP} 
V.V.Cheianov, M.Pustilnik, Phys.Rev.Lett. 100 (2008) 126403. 

\bibitem{G}
A.Imambekov, L.I.Glazman, Phys.Rev.Lett. 102 (2009) 126405. 

\bibitem{RMP}
A.Imambekov, T.L.Schmidt, L.I.Glazman, Rev.Mod.Phys. 84 (2012) 1253. 

\bibitem{S} 
A.Imambekov, L.I.Glazman, Science 323 (2009) 228. 

\bibitem{O}
A.A.Ovchinnikov, J.Stat.Mech. (2016) 063108. 

\bibitem{K1} 
N.Kitanine, K.K.Kozlowski, J.M.Maillet, N.A.Slavnov, V.Terras, 
J.Math.Phys. 50 (2009) 095209; J.Stat.Mech. (2011) P05028. 

\bibitem{K2} 
N.Kitanine, K.K.Kozlowski, J.M.Maillet, N.A.Slavnov, V.Terras, 
J.Stat.Mech. (2012) P09001. 

\bibitem{Korepin} 
V.E.Korepin, Teor.Mat.Fiz. 41 (1979) 169. 

\bibitem{LSM} 
E.Lieb, T.Schultz, D.Mattis, Ann.Phys. 16 (1961) 407. 

\bibitem{Shashi} 
A.Shashi, L.I.Glazman, J.S.Caux, A.Imambekov, 
Phys.Rev.B 84 (2011) 045408. 

\bibitem{Opla} 
A.A.Ovchinnikov, Phys.Lett. A375 (2011) 2694. 

\bibitem{OFF}
A.A.Ovchinnikov, J.Phys.:Condens.Matter 16 (2003) 3147. 
     
\bibitem{YY}
C.N.Yang, C.P.Yang, Phys.Rev. 150 (1966) 321. 

\bibitem{O19}
A.A.Ovchinnikov, J.Stat.Mech. (2019) 093103.

\bibitem{PWA1}
R.G.Pereira, S.R.White, I.Affleck, Phys.Rev.B 79 (2009) 165113.  


\end{thebibliography}
\end{document}